\documentclass[twocolumn,twoside,slac_two]{revtex4}
\usepackage{graphicx}
\usepackage{fancyhdr}
\begin{document}

\title{Strategies for the Detection of Gamma Rays from Dark Matter Annihilation Towards the Galactic Center Region with the High Energy Stereoscopic System}

%

\author{G. Spengler and U. Schwanke}
\affiliation{Humboldt University Berlin, Newtonstr. 15, D 12489 Berlin, Germany}

\begin{abstract}
The central region of the dark matter halo of the Milky Way is a promising target for a search for a particle dark matter annihilation signal. The H.E.S.S. Collaboration has published a search for a photon flux originating from dark matter 
particles annihilating in the galactic center region. No significant excess was observed and upper limits on the velocity averaged dark matter self annihilation cross section were derived. The limits exclude the self annihilation of dark matter particles in the 
$\sim 1$ TeV to $\sim 4$ TeV mass range with a velocity averaged annihilation cross section larger than $\sim 3\cdot 10^{-25}\:\mathrm{cm^3/s}$, i.e. about one order of magnitude above the prediction for a thermal relic dark matter particle. A detailed and 
realistic Monte Carlo study of new strategies for the search for a particle dark matter annihilation signal from the Milky 
Way dark matter halo with the High Energy Stereoscopic System is presented and the sensitivity of different 
experimental approaches is compared.

\end{abstract}

\maketitle

\thispagestyle{fancy}


\section{Particle Dark Matter and the Milky Way Dark Matter Halo}
The dynamics within the Milky Way points towards the existence of a large amount of invisible, i.e. dark, matter in the Milky 
Way. Many extensions of the standard model of particle physics naturally predict the existence of yet 
unobserved weakly interacting massive particles (WIMPs) that do not directly couple to light but only to the 
weak sector of the standard model. 
It is thus inviting to explain the failure to understand the dynamics within the Milky Way based on the visible distribution 
of matter by the additional abundance of WIMPs. The existence of WIMPs is also in agreement with cosmological 
structure formation arguments and precision measurements of the dark matter density in the universe.\\
In many models, WIMPs are expected to be stable on cosmological timescales but have the 
ability to annihilate into standard model 
particles. In the minimal supersymmetric extension of the standard model of particle physics with conserved R-parity, 
the neutralino, i.e. the superposition of supersymmetric partners to the neutral electroweak gauge bosons, is 
a Majorana particle dark matter candidate and is thus even able to self annihilate. Depending on the annihilation channel, 
the standard model annihilation products lead in part 
to a continuous photon spectrum. The photon energies can, depending on the particle dark matter mass, be in the 
Very High Energy (VHE, $E>100$ GeV) $\gamma$-ray regime. The number of expected particle dark matter annihilation events 
in a given volume is proportional to the dark matter density squared. Large dark matter densities are thus promising for 
a search for dark matter self annihilation events. 
Recent large scale N-body simulations of gravitationally interacting pseudo particles predict that the dark matter 
density distribution at the spatial scale of galaxies can be parametrized by simple functions and few parameters 
(see f.i. \cite{pieri}). 
\begin{figure}
\includegraphics[width=65mm]{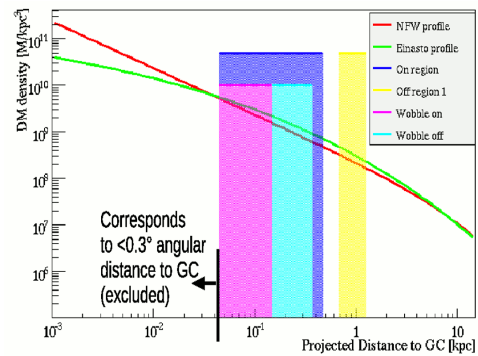}%
\caption{Two parametrizations of the dark matter density distribution as 
a function of the distance to the galactic center 
resulting from large scale N-body computer simulations. The two distributions correspond to the best fit to the dark matter 
distribution in two different Milky Way mass scale N-body computer simulations as obtained in \cite{pieri}. The dark matter 
density distributions agree for distances larger than 
$\sim 45$ pc within a factor of two. 
Both distributions are normalized to a dark matter density 
of $0.3\: \mathrm{GeV/cm^3}$ at a distance of $8.5$ kpc to the galactic center, i.e. the radial distance of the sun to the 
galactic center.}
\label{density}
\end{figure}
Figure \ref{density} shows the dark matter density distributions as recently obtained in two different Milky Way mass scale 
N-body computer simulations (see \cite{pieri}). Both distributions are normalized such that the density 
at the position of the sun agrees with the value of $0.3\: \mathrm{GeV/cm^3}$ derived from different dynamical constraints. 
A central result of N-body simulations involving only dark matter particles is that the dark matter density 
is in general steeply increasing towards the center of galaxies. The agreement between the resulting dark matter density 
distributions of different Milky Way mass scale N-body simulations is at the level of a factor of two for distances larger 
than $\sim 45$ pc from the center of the galaxy as apparent in Fig. \ref{density}. For more information on dark matter 
from an astrophysical and particle physics point of view see \cite{bertone} and references therein.
\section{The High Energy Stereoscopic System}
The High Energy Stereoscopic System (H.E.S.S.) is an array of Imaging Atmospheric Cherenkov Telescopes (IACTs) located in the 
Khomas Highland of Namibia at an altitude of $\sim 1800$ m above sea level. This study considers the H.E.S.S. I array 
of four identical $\sim 100\:\mathrm{m^2}$ mirror area IACTs operating in a square formation and optimized for the 
detection of $\sim 100$ GeV to $\sim 100$ TeV $\gamma$-rays from astrophysical sources. 
The instrumental field of view (FoV) of H.E.S.S. I is $\sim 5^\circ$ in diameter. The energy 
and angular resolution for the reconstruction of $\gamma$-rays is $\sim 20\%$ and $\sim 0.1^\circ$ respectivly. 
For more technical information on H.E.S.S. I see \cite{hessCrab} and references therein.  
\subsection{Background Subtraction and H.E.S.S. I Point Source Sensitivity}
The detection of $\gamma$-rays is complicated due to the necessity to distinguish $\gamma$-ray signal events from background 
events from Cosmic Rays. H.E.S.S. I performs a 
discrimination between Cosmic Ray and $\gamma$-ray events in multiple steps 
at trigger (\cite{hesstrigger}) and subsequent data analysis (\cite{hessCrab}) level. However, a perfect discrimination is not 
possible on an event by event basis and enforces the application of a background subtraction algorithm. Frequently used 
and well tested background subtraction algorithm rely on the definition of a signal and a background region 
in the same FoV (see \cite{berge} for an overview). 
The background region has to be constructed such that the instrumental acceptance 
in the signal and background region are 
equal. The number of expected background events in the signal region is then measured under the assumption that the 
Cosmic Ray TeV flux is isotropic. This background subtraction approach is well suited for the investigation of signal 
regions which are much smaller than the FoV and leads typically to the ability to 
detect astrophysical $\gamma$-ray point sources with a photon flux of $\sim 1\%$ of the Crab nebula 
$\gamma$-ray flux (\cite{hessCrab}) at a statistical 
significance of $5$ standard deviations within an observation time of $\sim 25$ h under optimal observation conditions for 
H.E.S.S. I. For $\gamma$-ray sources that are comparable in size or larger than the FoV of H.E.S.S. I the background region 
can however not be constructed in this way and alternative methods have to be applied.
\section{The Rotated Pixel Method}
\begin{figure}
\includegraphics[width=65mm]{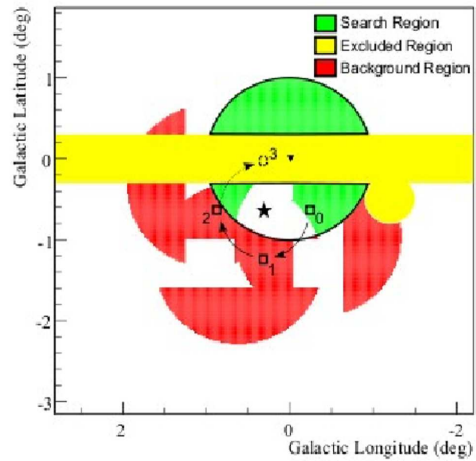}%
\caption{Signal (green) and background (red) region constructed with the rotated pixel background algorithm in galactic coordinates. Each pixel 
in the signal region is rotated out of the signal region at constant angular distance to the pointing direction. Known 
astrophysical $\gamma$-ray sources and the galactic plane are excluded (yellow).}
\label{refpix}
\end{figure}
The main technical problem in the search for a $\gamma$-ray signal from the 
galactic dark matter halo is that the halo extension is much larger than the H.E.S.S. I FoV. 
Traditional background subtraction algorithm that rely on the signal plus background and a background measurement in 
the same FoV can thus not easily be applied. 
The H.E.S.S. collaboration has recently published (\cite{daniil}) 
a search for a $\gamma$-ray signal from self annihilating WIMPs in the 
center of the Milky Way dark matter halo which uses an adaptation of the reflected region background subtraction algorithm (see 
\cite{berge}). The method employed relies on the steepness of the predicted dark matter 
density profile in the vicinity of the galactic center and the assumption of the H.E.S.S. I instrumental acceptance to be 
rotationally symmetric around the pointing position. A signal region of $1^\circ$ angular radius around the 
galactic center is 
defined and subdivided into many pixel each of which is much smaller than the angular resolution of H.E.S.S. I. For a given 
observation run pointing towards a constant direction in celestial coordinates, each pixel in 
the overlap of the signal region and the FoV is then rotated out of the signal region at constant angular distance 
to the pointing position to construct the background region for a run. 
Known astrophysical $\gamma$-ray sources, the galactic plane ($|b|<0.3^\circ$) and signal 
pixels for which no background pixel can be constructed for geometric reasons are excluded from the analysis.  
Figure \ref{density} shows the projected distances to the galactic center covered in the signal (magenta, tagged 'wobble on') 
and background (turquoise, tagged 'wobble off') 
region. The construction of the background region ensures that the signal region is always closer in angular distance 
to the galactic center than the background region and the predicted average dark matter density in the signal region 
is thus larger than in the background region. The background subtraction method 
relies thus not on a measurement of signal plus background and background in the same FoV but on a measurement of 
signal plus background and 'less signal' plus background in the same FoV. A search for a particle dark matter 
self annihilation signal is possible by comparing the number of events detected in the signal region with 
the number of events detected in the background region. The H.E.S.S. collaboration did not detect a significant signal 
in a rotated pixel analysis of $112$ h livetime of H.E.S.S. I observations of the galactic center region. The 
velocity averaged self annihilation cross section $<\sigma v>$ was in consequence constrained to be smaller than 
$\sim 3\cdot 10^{-25}\:\mathrm{cm^3/s}$ for WIMP masses around $1$ TeV (\cite{daniil}). The derived limits on $<\sigma v>$ 
are currently the most constraining in the TeV WIMP mass range. 
\subsection{Limitations}
The application of the rotated pixel background subtraction algorithm limits the angular extension of the signal region by 
the demand to fit signal and background region into one H.E.S.S. I FoV. A larger signal region could, depending on the 
signal to background ratio, increase the sensitivity of H.E.S.S. I to a WIMP self annihilation $\gamma$-ray flux from the 
galactic dark matter halo. Without the constraint due to the rotated pixel background subtraction, the maximal signal region 
could be increased to be complete H.E.S.S. I FoV.\\ 
The maximal distance of the background region to the galactic center is also limited by the applied 
background subtraction technique. If the background 
region can be constructed much further away from the galactic center than possible with the rotated pixel technique, the 
expected $\gamma$-ray flux due to WIMP self annihilation in the background region would be smaller and the sensitivity would 
thus improve.\\
In the following, two alternative background subtraction techniques that rely on special observation strategies are 
described and it is investigated whether an increase in sensitivity to a WIMP self annihilation $\gamma$-ray flux from the 
Milky Way dark matter halo is possible.  
\section{On/Off Background Subtraction}
\begin{figure}
\includegraphics[width=65mm]{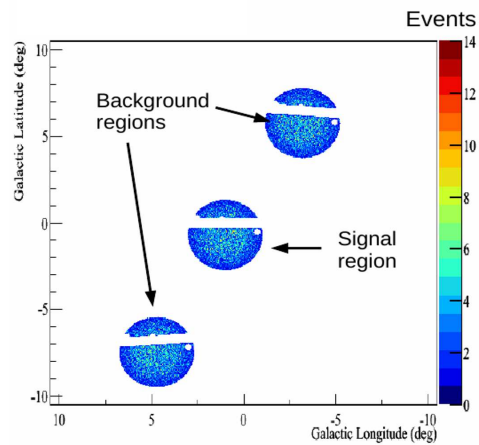}%
\caption{Simulated H.E.S.S. I events for the observation of a signal region with the full H.E.S.S. I FoV and two 
background regions in galactic coordinates. 
The pointing positions for the background regions have a symmetric offset in right ascension to the 
pointing position of the signal region. Known $\gamma$-ray sources and the galactic plane are excluded from the analysis and 
all exclusion regions are mutually applied to all observation regions in the FoV system.}
\label{onoff}
\end{figure}
Figure \ref{onoff} shows simulated H.E.S.S. I $\gamma$-ray events in three different FoVs. H.E.S.S. I extragalactic 'off data', 
i.e. observation runs with no $\gamma$-ray source in the FoV, performed at a zenith angle of $\sim 20^\circ$ 
were used to obtain a realistic $\gamma$-ray candidate 
event rate after H.E.S.S. I standard $\gamma$-ray selection criteria (cuts). Additionally, the dependence of the $\gamma$-ray 
candidate event rate after cuts as a function of the offset to the observation position ('radial acceptance') 
and the energy dependence of the event rate after standard cuts together with the instrumental dead time 
as obtained from extragalactic off data are used to 
simulate realistic observations. Figure \ref{onoff} shows in the center the simulated signal region events in the vicinity 
of the galactic center distributed over 
a complete H.E.S.S. I FoV. The observation of two background regions whose observation positions have 
a symmetric offset of $\pm 35$ min in right ascension to the signal region observation position is simulated. 
Known $\gamma$-ray sources and the galactic plane ($|b|<0.3^\circ$) are excluded from the analysis and all exclusion regions 
are mutually applied to all observed FoVs respectively. For each simulated FoV, $50$ h observation livetime were generated.\\
The simulation models a peculiar observation strategy where always three H.E.S.S. I standard observation runs 
are taken consecutively such that the zenith and azimuth pointing angle range is always equal for a sequence of three runs. 
One observation sequence starts with the observation of a background region with 
$-35$ min offset in right ascension from the signal region pointing position and is 
performed with a standard length of $33$ min. 
The array pointing moves forward by $35$ min in right ascension immediately after the end of the first observation within 
a realistic run transition time of $2$ min. The second run of the sequence observes then the signal region for $33$ min. 
The third observation is again scheduled for $33$ min immediately after the second observation 
by moving the array pointing again by $35$ min forward in right ascension within $2$ min. Finally, the number of 
events after the application of standard $\gamma$-ray cuts in the signal and and the corresponding number of events 
in background region are compared. 
The observation strategy ensures as stated that the zenith and azimuth pointing angle 
range is always the same for every run 
within a run sequence. The strong dependence of the H.E.S.S. I acceptance on the pointing zenith and azimuth angle does 
in consequence not influence the calculation of a $\gamma$-ray excess as the dependence is the same for the signal and 
the background region. Repeated taking of three run sequences in this observation pattern can accumulate the simulated 
amount of data. The advantage is that obviously the signal region can be as large as the H.E.S.S. FoV (see also the 
blue region in fig. \ref{density}) and the background 
regions (see also the yellow region in fig. \ref{density}) 
are further away from the signal region with the result of less expected WIMP self annihilation $\gamma$-ray flux in the 
background region. This leads to the expectation that the sensitivity of the On/Off observations to a WIMP self annihilation 
$\gamma$-ray flux in the Milky Way halo is better than for the rotated pixel background subtraction technique. On the 
other hand, the method needs $2/3$ of the total observation time for the estimation of the expected number of background 
events in the signal region where the rotated pixel algorithm can use the complete observation time. It is a non trivial 
question which of the two competing factors is eventually dominant and therefore this realistic simulation is performed. 
\section{Driftscan Observations}
\begin{figure}
\includegraphics[width=65mm]{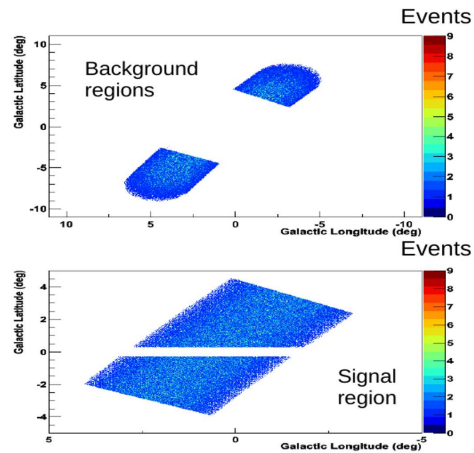}%
\caption{Simulated H.E.S.S. I events for the driftscan observation of the Milky Way dark matter halo in galactic coordinates.
Simulated is the repeated taking of data at constant zenith and azimuth pointing where the galactic center region 'drifts' 
through the FoV after approximately half of the observation time. Each run is subsequently divided into two parts, the signal 
region enclosing the galactic center region in the lower panel and the background region in the upper panel. Known $\gamma$-ray 
sources and the galactic plane ($|b|<0.3^\circ$) are excluded. Exlcuded regions are treated with a special method (see text) 
to guarantee the same instrumental acceptance in the signal and background region.}
\label{drift}
\end{figure}
Figure \ref{drift} shows the simulation of events recorded in driftscan observations of the galactic center region. 
Driftscan runs are scheduled at a constant zenith and azimuth angle array pointing. The celestial pointing at the beginning 
of a run is chosen such that the galactic center region 'drifts through' the FoV after 
approximately half of the $68$ min observation time. As for 
the simulation of On/Off events, realistic event rates for the observation at $\sim 20^\circ$ zenith angle 
after standard H.E.S.S. I $\gamma$-ray cuts have been used. Also the radial acceptance 
as well as the energy dependence of the events after $\gamma$-ray cuts and the instrumental dead time resemble real 
observation conditions. A total of $150$ h of observation time is simulated. The observed region is divided into many pixels 
each of which is much smaller than the H.E.S.S. I angular resolution.  Each run is divided in time into one signal 
region enclosing the galactic center and one background region. 
The division of a driftscan run into two halves is realized such that the 
product of the average time a pixel stays in the H.E.S.S. I FoV with the total solid angle of the region is the same 
for the signal and background region. As for the rotated pixel 
technique and the On/Off observations all known $\gamma$-ray sources and the galactic plane ($|b|<0.3^\circ$) are 
excluded from the analysis. Excluded pixel are shifted mutually in right ascension 
between signal and background region at constant declination and special care is taken to guarantee that the 
shifting of the exclusion regions does not imbalance the instrumental acceptance of signal and background region. 
Every pixel excluded in the signal region is 
shifted along right ascension and constant declination as close as possible to the edge of the background region to 
minimize the expected $\gamma$-ray flux from WIMP annihilation in the background region. Similarly, every pixel excluded 
in the background region is shifted as close as possible to the edge of the signal region to maximize the expected 
$\gamma$-ray flux from WIMP annihilation in the signal region.\\
The observation strategy and the subsequent data analysis ensure again, as for the On/Off and rotated pixel method, 
an equal instrumental acceptance in the signal and background region. 
Also similar to the On/Off method, the signal region is larger than for the rotated pixel method and the background region 
is further away from the galactic center than possible for the rotated pixel method. These effects should increase the 
sensitivity to a WIMP annihilation $\gamma$-ray flux from the galactic halo. On the other hand, the increase in 
sensitivity competes with the decrease in sensitivity due to observation time only used for background observations and 
the average expected signal flux per signal pixel is smaller than for the On/Off method because the signal FoV is larger. 
Again, it is non trivial to estimate which effect dominates and a realistic simulation that compares all three 
discussed methods under equal conditions is necessary. 
\section{Sensitivity Comparison}
\begin{figure}
\includegraphics[width=65mm]{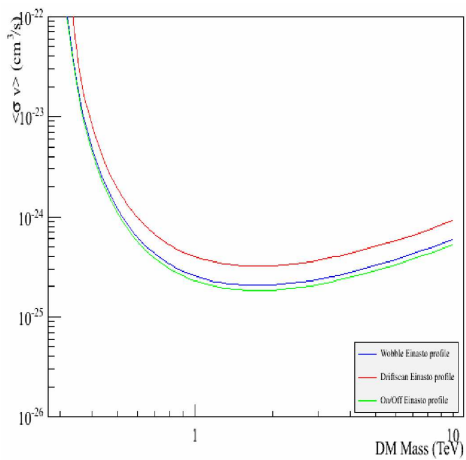}%
\caption{Sensitivity comparison between the rotated pixel method (blue), the On/Off method (green) and the driftscan 
method (red). Shown is the sensitivity (average $95\%$ C.L. upper limit) to the velocity averaged self annihilation 
cross section $<\sigma v>$ as a function of the assumed WIMP mass in case that no signal can be detected.}
\label{sensi}
\end{figure}
Figure \ref{sensi} shows the sensitivity of the rotated pixel, the On/Off and the driftscan method to the 
velocity averaged self annihilation cross section $<\sigma v>$ as a function of the WIMP mass. The Einasto parametrization 
for the dark matter density distribution of the Milky Way is assumed, parameters are taken from \cite{pieri}. The 
continuous $\gamma$-ray spectrum expected from the WIMP self annihilation is adopted from \cite{tasitsiomi}. 
For each investigated method, the On/Off, the driftscan and the rotated pixel method a total of $150$ h of observation 
time is simulated. No WIMP annihilation $\gamma$-ray signal is simulated but only background events such that no significant 
$\gamma$-ray excess is resulting in all cases and an upper limit on $<\sigma v>$ can be derived which serves as a 
sensitivity estimation for the method under consideration. The upper limit on $<\sigma v>$ itself is calculated at 
$95\%$ confidence level using H.E.S.S. I effective areas and energy threshold lookups for standard Hillas cuts (\cite{hessCrab}) 
and with a method that leads to results that are compatible with the method employed in \cite{daniil} 
under equal conditions.\\
Figure \ref{sensi} shows that the sensitivity to $<\sigma v>$ of each method investigated is comparable in the order of 
magnitude. The On/Off method is slightly more sensitive ($\sim 20\%$ at $1$ TeV WIMP mass) than the rotated pixel method. 
The driftscan method is less sensitive ($\sim$ a factor of $2$ at $1$ TeV WIMP mass) 
than the rotated pixel method and the On/Off method. However, the sensitivity of the driftscan method can be increased 
compared to the result shown in Fig. \ref{sensi} when the size of the driftscan background region is increased at the cost 
of the size of the signal region. In an optimized signal to background region size case the sensitivity of the driftscan 
method agrees with the sensitivity of the On/Off and rotated pixel method within $\sim 30\%$.\\
Overall it can be stated that the increase in sensitivity of background subtraction methods alternative to the rotated 
pixel method is not as large as could be expected. The reason for this is the large amount of observation 
time that the alternative background subtraction methods need only for background observations. On the other hand, 
it is obvious that the alternative background subtraction methods can be sensitive to a WIMP annihilation $\gamma$-ray 
flux even in the case when the dark matter density profile is for some reason nearly constant in the vicinity of the 
Milky Way halo center but only decreasing at large distances to the galactic center. Additionally, the sensitivity of 
the driftscan and On/Off method is larger than the sensitivity of the rotated pixel method if the center of the 
Milky Way dark matter halo is not spatially coincident with the galactic center but offset by $\sim 1^\circ-2^\circ$ as 
possibly indicated in \cite{dmoffset}. In conclusion it is (despite of the sensitivity to the discussed benchmark 
model with an Einasto dark matter density profile being comparable to the rotated pixel technique) 
of interest to record at least a limited amount of observational data 
with the discussed observation strategies.
%



\bigskip 

\end{document}